\documentclass[preprint,12pt]{elsarticle}
\usepackage{lineno,hyperref}
\usepackage{graphics}
\usepackage{comment}
\usepackage{breakurl}
\modulolinenumbers[5]

\journal{Journal of \LaTeX\ Templates}

\begin{document}

\begin{frontmatter}

\title{Performance evaluation of GAGG(Ce)/LFS scintillator + MPPC array readout with ASIC}

\author[1]{Daiki Nobashi\corref{cor1}}
\ead{nobashi.daiki@isee.nagoya-u.ac.jp}
\author[1]{Kazutaka Yamaoka}
\author[1]{Hiroyasu Tajima}
\author[2]{Kazuya Ito}

\cortext[cor1]{Corresponding author}

\address[1]{Institute for Space-Earth Environmental Research (ISEE), Nagoya University, Furo-cho, Chikusa-ku, Nagoya 464-8601, Japan}
\address[2]{Electronic Information Technology Office, Technical center, Nagoya University, Furo-cho, Chikusa-ku, Nagoya 464-8602, Japan}

\begin{abstract}

We constructed a gamma-ray detector by combining two types of scintillator array detectors with an MPPC array and evaluated the spectral performance by reading out the signals from the MPPC with a low-power integrated circuit (ASIC) manufactured by IDEAS in Norway. One of the two types of scintillators is a GAGG(Ce) (Ce-doped $ \rm{Gd_3Al_2Ga_3O_{12}}$) scintillator, and the other is an LFS scintillator. The scintillator array is 2.5 cm $\times$ 2.5 cm in size and is coated with $ \rm{BaSO_4}$-based white paint for GAGG(Ce) and an enhanced specular reflector (ESR) for LFS except for the side optically coupled to the MPPC. The spectra derived from the array are affected by the MPPC photon saturations and light leakage from the adjacent pixels, and we carefully corrected for both effects in our data analysis. The energy resolution of 662 keV at 20 $^\circ$C is 6.10$\pm$0.04\% for the GAGG(Ce) scintillator array and 8.57$\pm$0.15\% for the LFS scintillator array, this is equivalent to the typical energy resolution found in the references. The energy resolution depends on the temperature: the energy resolution improves as the temperature decreases. We found that the contribution of thermal noise from the MPPCs to the energy resolution is negligible within the range of --20 to 40 $^\circ$C, and the energy resolution is mainly determined by the light yield of the crystals. 

\end{abstract}

\end{frontmatter}

%\linenumbers

\section{Introduction}

The soft gamma-ray within the 100 keV to 1 MeV or higher range is an unexplored wavelength in astrophysics, although there are many scientifically important targets, e.g., non-thermal radiation related to particle accelerations in a supernovae explosion and black holes, and a nuclear gamma-ray line emission from $^{44}\rm{Ti}$ and $^{26}\rm{Al}$ and a 511 keV line gamma-ray from electron-positron annihilation. 
This is because it is difficult to focus soft gamma-rays on detectors utilizing optics and measure the incident energy because Compton scattering is a dominant physical process used in detection techniques. 
Hence, the detector suffers from a rather high instrumental background compared to signals from celestial sources, reducing the signal-to-noise ratio. 
A Compton camera is mainly used in soft gamma-rays, which measures the incident direction and energy by using events that are scattered by the scatterer and then photoelectrically absorbed by the absorber.  
Actually, no astrophysical observations using a Compton camera have been made since COMPTEL \cite{COMPTEL}, a large Compton camera using liquid scintillators and NaI(TI) scintillators, onboard the Compton Gamma-Ray Observatory (CGRO), was deorbited in 2000. 

Recently, with the advancement of instrumentation technologies such as integrated circuits and semiconductor technology, a Compton camera using semiconductor detectors have the potential tools to achieve much higher sensitivity than COMPTEL. 
One of them is the Soft Gamma-ray Detector (SGD) onboard the Hitomi satellite, which uses the a Si semiconductor sensor as a scatterer and a CdTe semiconductor sensor as an absorber and has a sensitivity of 60 to 600 keV \cite{ASTRO-H}.
If this sensitivity is extended up to 1 MeV or more with only semiconductor sensors, CdTe semiconductor layers (0.75 mm $\times$ 8 layers, total 6 mm) would need to be as thick as a few centimeters, which is more difficult both technically and budgetary. 
As an absorber for MeV regions, the easiest way is to develop a hybrid Compton camera combined with an inorganic scintillator, which has a high stopping power and can be easily processed with a thickness of several centimeters. 
During the 21st century, new scintillators with a high performance in terms of light yield, time response, and stopping power have been developed, which exceeds the performance of conventional NaI(Tl) crystals. 

In this paper, we evaluate a 4$\times$4 (6 $\rm{mm}$ cubic for each) array using two types of scintillators, cerium-doped gadolinium aluminum gallium garnet (GAGG(Ce)) and lutetium fine silicate (LFS), a readout with a multi-pixel photon counter (MPPC) array provided by Hamamatsu Photonics K.K. and an application specific integration circuit (ASIC) IDE3380 \cite{IDEAS_ASIC} provided by Integrated Detector Electronics AS Co. (IDEAS) for future Compton camera applications in space.
These two scintillators were newly developed in the 2000s and show an excellent performance in terms of a high light yield, fast time response, and high stopping power. 
In Section 2, we describe the experimental setup. In Section 3, we report on the analysis and results of the spectral performance and temperature dependence of the detector. 
Finally, we provide some concluding remarks.  

% Setup
\section{Experimental Setup}
\subsection{Inorganic scintillator arrays and Multi-Pixel Photon Counter (MPPC)}

\begin{figure}[htb]
\begin{center}
\includegraphics[width=0.7\linewidth,bb=0 0 354 200]{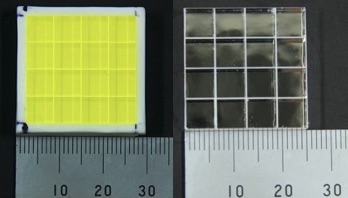}
\caption{An image of GAGG(Ce) scintillator array covered with BaSO$_4$ (left) and LFS scintillator array covered with ESR (right).}\label{fig1}
\end{center}
\end{figure}

We constructed two types of 4$\times$4 MPPC+scintillator array for the readout test using the ASIC IDE3380. One is the GAGG(Ce) (Cerium-doped $\rm{Gd_3 Al_2 Ga_3 O_{12}}$) scintillator manufactured by Furukawa Denshi Co., Ltd., and the other is an LFS ($\rm{Ce_xLu_{2+2y-x-z}A_zSi_{1-y}O_{5+y}}$: A= Gd, Sc, Y, La, Eu, or Tb) scintillator manufactured by Zecotek Photonics, Inc. 
Both crystals have a higher stopping power than those of NaI(Tl) and CsI(Tl). 
Table 1 shows the characteristics for both scintillators in comparison with other crystals. 
The scintillator array consists of cubic crystals with a side length of 6 mm arranged in a 4$\times$4 matrix with each gap of 0.2 mm. The overall size is 25 mm $\times$ 25 mm. 
All six faces of crystals are polished, and different reflectors are used in the gap and surrounding: BaSO$_4$-based white reflector for the GAGG(Ce) array and a 3M Enhanced Specular Reflector (ESR) for the LFS array (see Figure \ref{fig1} for images).  
The array size is completely matched to the sensitive area of the Hamamatsu MPPC 4 $\times$ 4 array S14161-6050HS-04. 
This MPPC array is newly developed for the Positron-Emission Tomography (PET) application and has a higher photon detection efficiency (PDE) and lower operation voltage than those of the S13360 series. The PDE is 42\% and 50\% for the GAGG(Ce) peak of 520 nm and LFS peak of 425 nm, respectively. 
This surface-mount type MPPC array was mounted on the electric board and RC low-pass filters with $R$ = 1 k$\Omega$ and $C$ = 1 $\mu$F were put near the array to reduce high frequency noise from the power supply.
The scintillator array was attached to the MPPC array with optical grease EJ-550 (ELJEN). 
\setlength\intextsep{2pt}
\begin{table*}[h]
\begin{center}
\caption[]{Characteristic of typical scintillators}
\small
\begin{center}
	\begin{tabular}{|c|c|c|c|c|c|}
		\hline
		& GAGG(Ce) & LFS & NaI(Tl) & CsI(Tl) & BGO \\
		\hline
		Density (g $ \rm{cm^{-3}}$) & 6.63&7.35&3.67&4.53&7.13 \\
		Light yield (photons $ \rm{MeV^{-1}}$) & 60000& 36000--38000 & 45000 & 56000 & 8000 \\
		Effective atomic number & 54&64&51&54&74\\
		Decay time (ns) &88&33&230&1050&300 \\
		Peak emission wavelength (nm) &520&425&415&550&480 \\
		Deliquescence & no & no & yes & yes& no \\
		\hline
	\end{tabular}
\end{center}
\end{center}
\end{table*}

\setlength\intextsep{3pt}
\subsection{Application Specific Integrated Circuit (ASIC)}

For the readout of the MPPC+GAGG(Ce)/LFS array, we used the IDEAS evaluation system, which consists of an ASIC IDE3380 test board, a so-called SIPHRA TG, and the GALAO board. 
The ASIC IDE3380 can process signals from 16-channel sensors under an extremely low-power consumption (less than 2 mW per channel).  
The ASIC has current divider circuits (current mode input stage (CMIS)), 
 charge-sensitive amplifiers (charge integrator (CI)), shaping amplifiers (SHA), and trigger generation circuits. 
 A sample-hold analog-to-digital converter (ADC) is also implemented in the ASIC, and it can digitize the pulse height information at the pulse peak timing. 
It is possible to program the gain, shaping time, and threshold voltage for the input signal for each of the 16 channels. 
There are two types of gain for each of the 16 channels in the CMIS and CI. The CMIS can downscale the current input from the MPPC mainly to four stages (1/10, 1/100, 1/200, and 1/400), and the CI then integrates the current from the CMIS in four steps (1 V/30 pico-coulomb (pC), 1 V/27.75 pC, 1 V/3 pC, and 1 V/0.75 pC). 
By adjusting these two gains, we can determine the dynamic range of the sensors. 
The shaping time in the SHA can be changed to four different values (200, 400, 800, and 1600 ns) through the configuration register. 
The output data consist of the time of the event, the trigger type, the triggered channels, and 12-bit ADC data. 
The GALAO board, equipped with a field programmable gate array (FPGA), is the interface between a PC and an ASIC, and mainly reads serialized data and controls the ASIC. 
The ASIC settings are programmed from the PC through the GALAO board using the benchmark software provided by IDEAS. 
The deadtime for processing the signals from 16 channels is approximately 209.6 $\mu$s. 

\subsection{Measurement Setup}

\begin{figure}[h]
\begin{center}
\includegraphics[width=0.8\linewidth,bb=0 0 935 439]{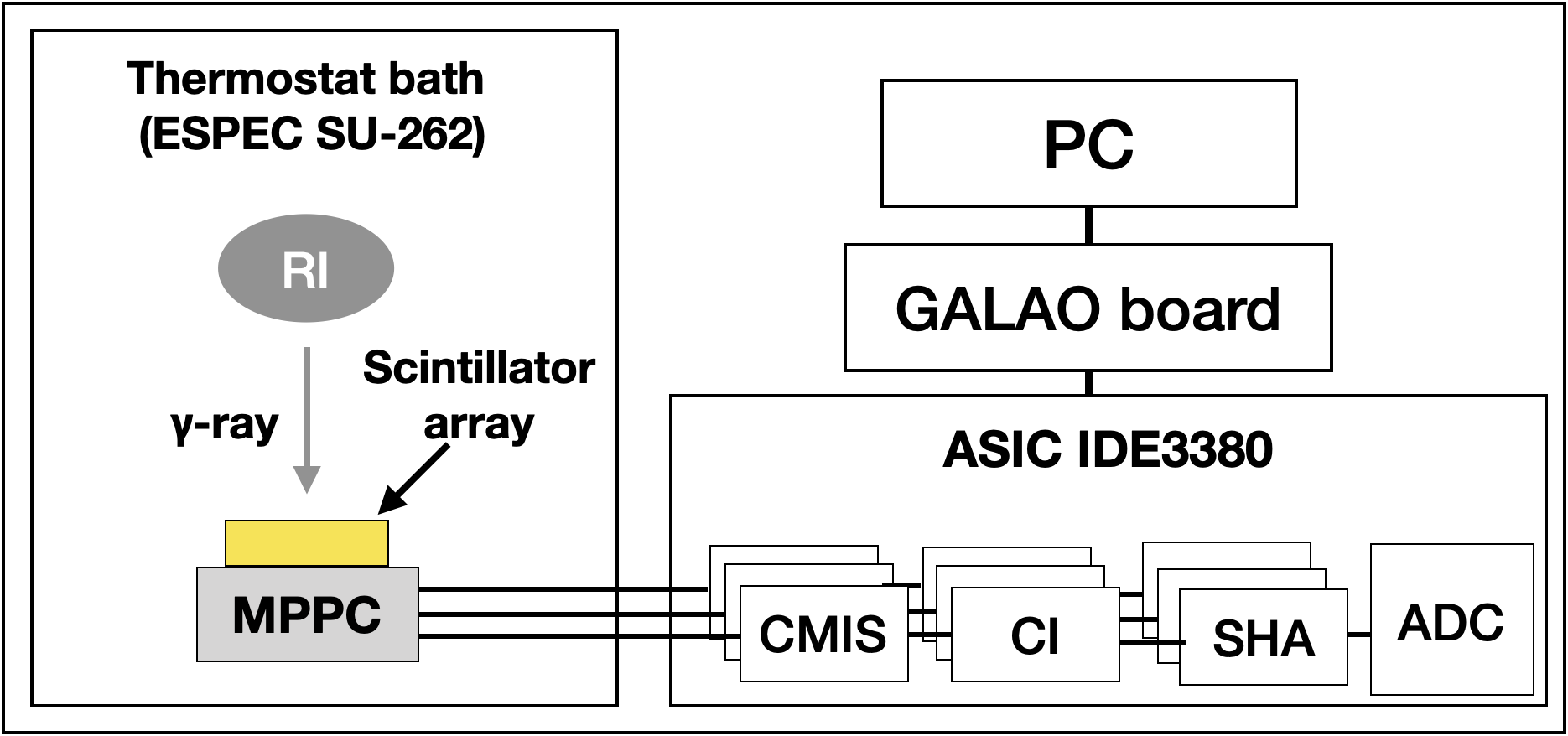}
\caption{Experimental setup for the measurement of each of the two scintillator arrays (GAGG(Ce) and LFS).
The temperature of the scintillator array controlled by the thermostat bath and the sensor signals from the scintillator array with the MPPC array are read out using an ASIC IDE3380. Data acquisition is conducted using the GALAO board and PC.}
\label{fig2}
\end{center}
\end{figure}

The overall system setup is shown in Figure \ref{fig2}. 
Scintillation detectors consisting of an MPPC array and two scintillator arrays were placed in a thermostat bath (ESPEC SU-262), and the temperature in the bath was set to the four different temperatures (--20, 0, 20, and 40 $^\circ$C) to evaluate the performance. 
A $^{137}$Cs source with an intensity of 40 kBq was used to irradiate the arrays with 662 keV gamma-rays.
To adjust the ASIC negative charge input, the negative bias voltage was provided to the anode side of the MPPC using a KEITHLEY 6487.
The gain of the MPPC significantly changes with the temperature. 
To measure the intrinsic temperature dependence of the scintillators, the operation voltage of MPPC array was carefully changed such that an over-voltage of 3.0 V was applied at any temperature. 
The shaping time, CMIS gain, and CI gain are set to 800 ns, 1/200, and 1 V/27.75 pC, respectively. 
The input charge can be measurable at up to $\sim$10000 pC in this setting.

% Result
\section{Analysis \& Results}

We have analyzed 4$\times$4 array data at several stages consisting of basic corrections (pedestal subtraction, conversion of ADC channel into a charge, 
background subtraction, and deadtime correction), single event selection using the center of gravity (COG) method, and corrections of the MPPC saturation and light leakage effect.  
We describe each of these in the following subsections. 

\subsection{Basic corrections}

The pedestal and linearity between the input charge and ADC channels are measurable using a test pulse function in the GALAO board.
The pedestal value is distributed within 140 to 260 ADC channels in the 16-channel array, and is always subtracted during the analysis.  
The linearity curve was fitted using fourth-order polynomial function and was converted from an ADC channel to a charge in units of pC. 

We found an extremely high intrinsic background when reading out the LFS scintillator array,
which was caused by $^{176}$Lu contained in the LFS scintillator. Here,
$^{176}$Lu predominantly emits two gamma-rays with an energy of 201.83 and 
 306.78 keV as well as beta-rays \cite{LFS_background}. The background rate at above 10 keV was only approximately 14 Hz for the GAGG(Ce) scintillator array, and was 1050 Hz for the LFS scintillator array. Hence, particularly for the LFS array, we found that the background was not negligible for the relatively weak $^{137}$Cs source we used, and the deadtime effect should also be considered for analysis. 
\begin{figure*}[htb]
	\begin{minipage}{0.5\hsize}
		\begin{center}
		\includegraphics[width=0.68\linewidth]{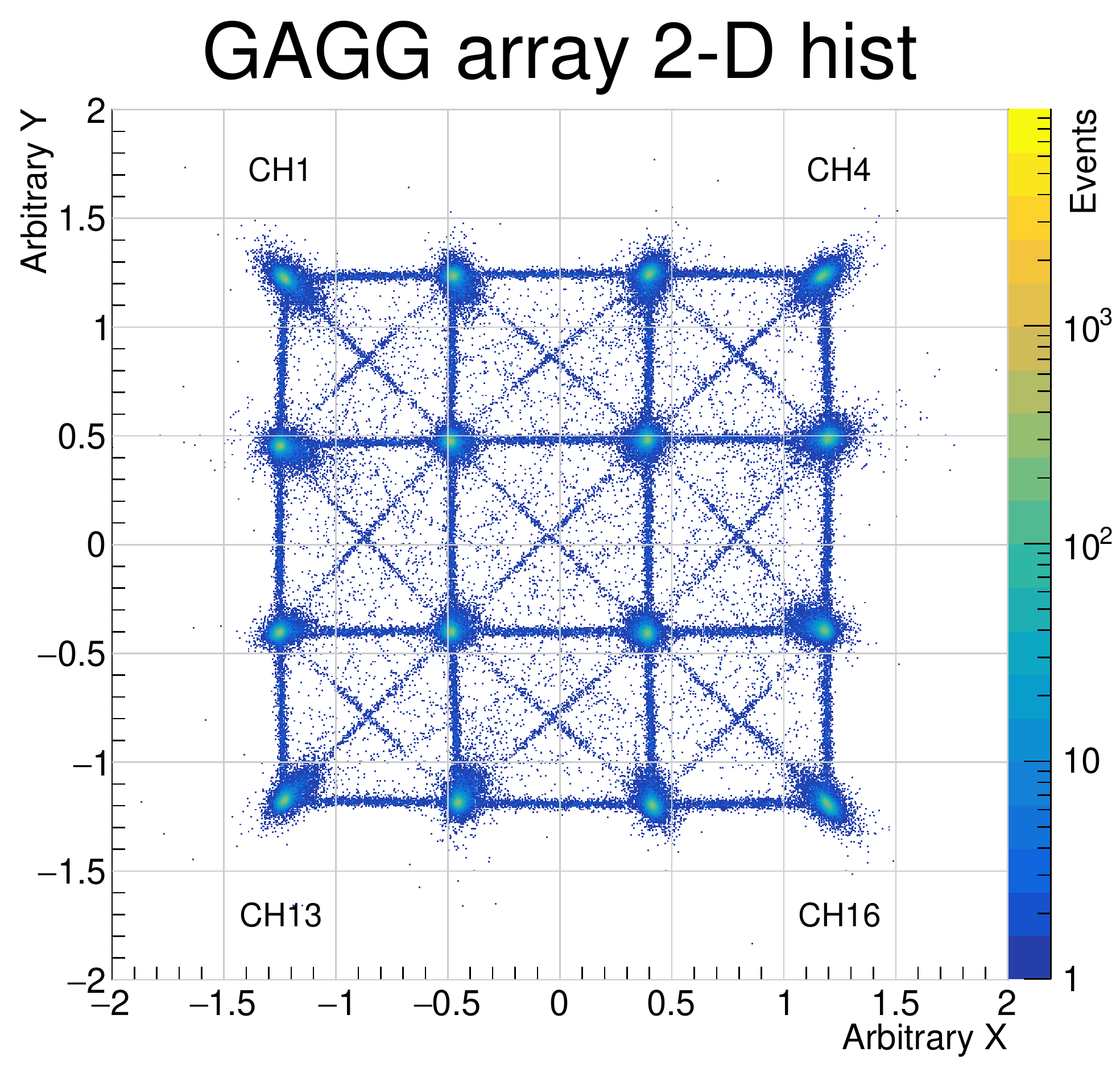}
		\end{center}
	\end{minipage}
	\begin{minipage}{0.5\hsize}
		\begin{center}
		\includegraphics[width=0.68\linewidth]{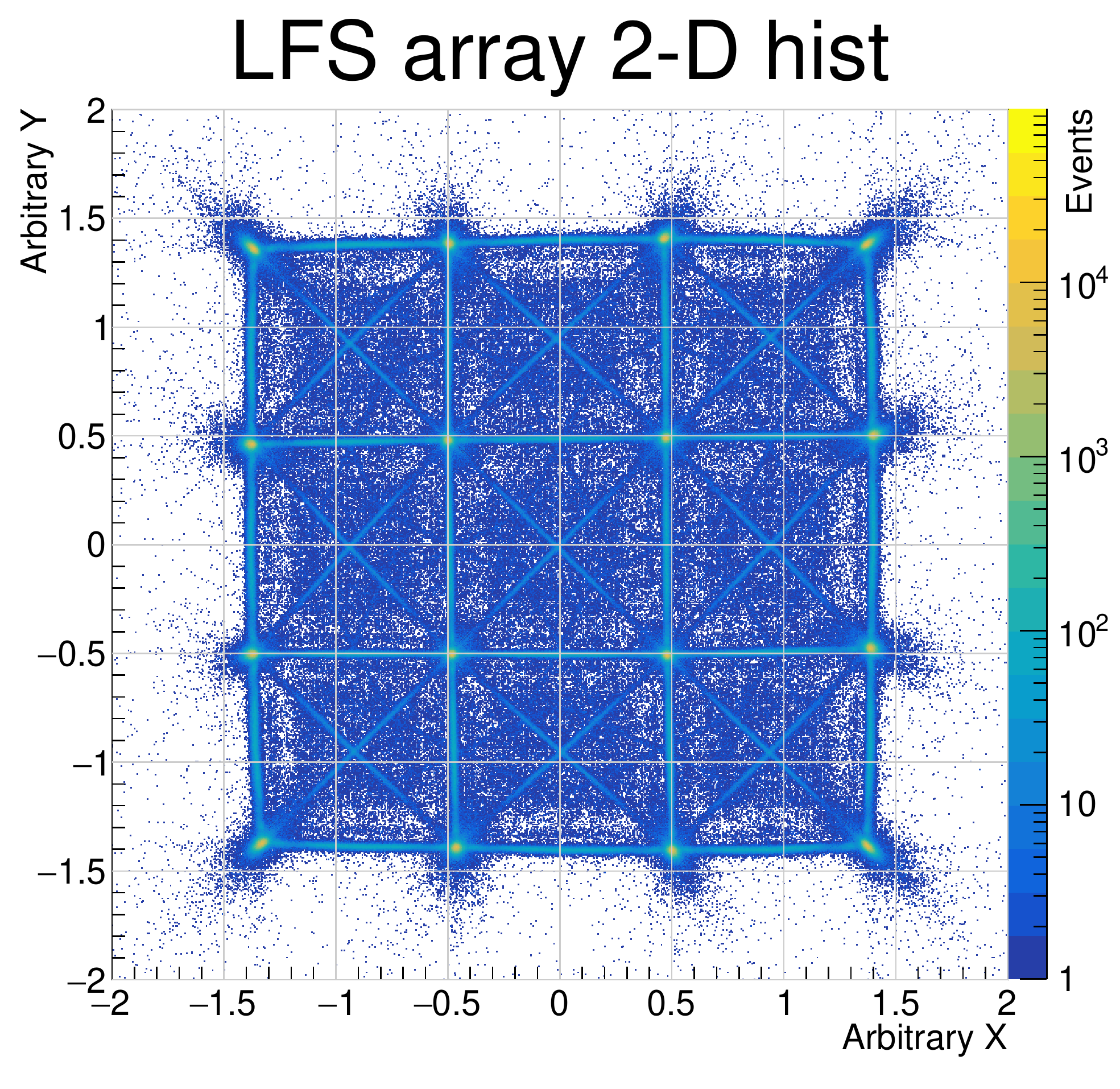}
		\end{center}
	\end{minipage}
	\caption{The center of gravity (COG \cite{center_of_gravity}) distributions of the events obtained when $^{137}$Cs was irradiated onto the GAGG(Ce) (left) and LFS (right) arrays. This is measured at 20 $^{\circ}$C, and is not background-subtracted.}
	\label{fig_cog}
\end{figure*}

\subsection{Single Hit Event Selection by Center of Gravity (COG) Method}

To obtain the position information of each gamma-ray event in the array, the center of gravity ($x_{\rm G}$,$y_{\rm G}$)
was calculated using the following equations.
\begin{equation}\label{eq1}
x_{\rm G}=\frac{\sum_{i=1}^{16} ADC_{i} \cdot x_{i}}{\sum_{i=1}^{16} ADC_{i}}, y_{\rm G}=\frac{\sum_{i=1}^{16} ADC_{i} \cdot y_{i}}{\sum_{i=1}^{16} ADC_{i}},
\end{equation}
where the $ADC_{i}$ is the ADC value for each CH ($i$=1, 2,...16 from upper-left to right-lower in Fig.\ref{fig_cog}), and $x_{i}$ and $y_{i}$ are position information, which takes a value at --1.5, --0.5, +0.5 and +1.5.  
The COG distribution of the GAGG(Ce) and LFS array is shown in Figure \ref{fig_cog}. If gamma-rays are reacted in only one pixel and there is no light leakage to adjacent pixels, the COG coordinates should take discrete values of ($x_{i}$, $y_{i}$), where $x_{i}$,$y_{i}$=--1.5, --0.5, +0.5, and +1.5. 
The 4$\times$4 COG peaks show a single interaction event in each CH, e.g., the coordinate of COG (--1.5, +1.5) corresponds to single hit events in the CH1.
However, for the GAGG(Ce) array, the COG scale decreases by a factor of 0.8--0.9 compared with such values, probably due to the light leakage effect. In addition, we can also see many straight lines connected between the COG peaks, e.g., the horizontal line between CH1 and CH2, vertical line between CH1 and CH5, and diagonal line between CH1 and CH6, owing to Compton scattering events between two pixels. All other events except for the COG peaks 
 correspond to events interacted in more than one pixel, i.e., a multi-pixel hit event. 
To select only single-hit detection events, we extracted events around a peak in the COG distribution. 

\subsection{MPPC Saturation and Light Leakage Effect}

The GAGG(Ce) is a scintillator, which yields a high light output of 60,000 photons MeV$^{-1}$, which could result in saturation
 of MPPC fired pixels. The saturation effect on an MPPC coupled to scintillators is described by the following equation:
\begin{equation}\label{eq2}
N_{\rm mes}=N_{\rm max} \left\{1-\exp{ \left(-\frac{\epsilon k N_{\rm scin}}{N_{\rm max}}\right)}\right\},
\end{equation}
where $N_{\rm mes}$ is the number of MPPC fired pixels; $N_{\rm max}$ is the effective maximum pixel number determined by the number of MPPC 
 pixels, the decay time of the scintillator, and the recovery time for a fired pixel; $N_{\rm scin}$ is all photons created within the scintillator; $\epsilon$ is the PDE of the MPPC; and $k$ is the light collection efficiency. The photon number ($N$) can easily be converted to a charge ($Q$) measured in the ASIC using $Q=eG \epsilon N$, where $G$ is the MPPC gain of $\sim 10^6$ and $e$ is the elementary charge of the electrons.

Light leakage from adjacent pixels cannot be avoided because each crystal in the array is not completely separated by reflectors.
The GAGG(Ce) and LFS arrays utilize different reflectors with a 0.2-mm thickness among the crystals, i.e., BaSO$_4$ and ESR, and hence 
 we found that the GAGG(Ce) array has more light leakage than the LFS array. This effect is expected to be proportional to the light
 output produced in one pixel. If we assume a light leakage factor from the pixel $i$ where the scintillation light is produced 
 to the pixel $j$ as $l_{ij}$, the relation between charges $Q_i$ and $Q_j$ in the pixel $i$ and $j$ is given by the following: 
\begin{equation}\label{eq3}
Q_j = Q_{\rm max} \left\{ 1- \left( \frac{Q_{\rm max}-Q_i}{Q_{\rm max}} \right)^{l_{ij}} \right\}  
\end{equation}
where we assume the maximum charge ($Q_{\rm max}=eG \epsilon N_{\rm max}$), the PDE ($\epsilon$), and the MPPC gain ($G$) to be common in all pixels. 

\begin{figure}[h]
\begin{center}
\includegraphics[width=0.7\linewidth,bb=0 0 567 583]{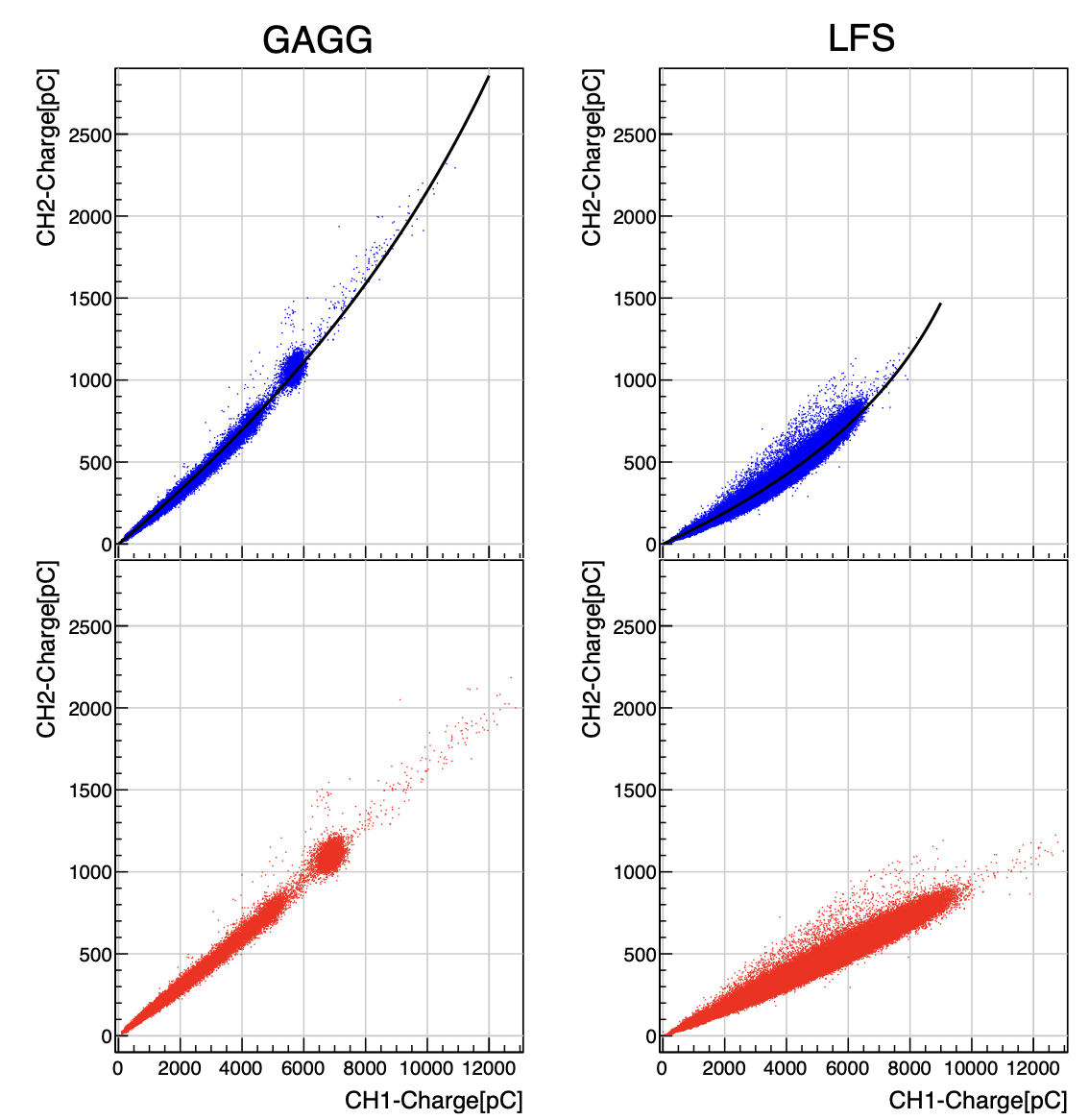}
\caption{Correlation distributions of between CH1 and CH2 before and after correction for the saturation effects.GAGG(Ce) (upper left) and LFS (upper right) are shown on the top and bottom before and after a correction of the saturation effects, respectively. The black curve in the upper panel shows the best-fit function of Equation (3), the parameters of which are given in Section 3.3. }\label{fig_correlation}
\end{center}
\end{figure}

Figure \ref{fig_correlation} shows a scattered plot between charges in CH1 and CH2 between GAGG(Ce) and LFS. 
A simple light leakage effect should follow a linear correlation between two pixels, but both correlations show a clear non-linear correlation with a curvature. 
We fitted data with Equation (\ref{eq3}) using two free parameters $Q_{\rm max}$ and $l_{ij}$, and the fit is reasonable as indicated by the best-fit function shown by the black curve in the upper panel of Figure \ref{fig_correlation}.
In this case, the best-fit parameters $\left(Q_{\rm{max}},\ l_{\rm 12}\right)$ are $\left(17912,\ 0.1567\right)$ for GAGG(Ce) and $\left(11314,\ 0.0877\right)$ for LFS, respectively.
The LFS scintillator has a lower light yield, but a faster decay time than those of GAGG(Ce).
The combined effect of the two parameters is considered to reduce the saturation level $Q_{\rm{max}}$ more than that of GAGG(Ce).
The charge $Q_{\rm real}$(=$eG \epsilon k N_{\rm scin}$) corrected for the saturation effect is given by Equation (\ref{eq4}), and shown by red in lower panel of Figure \ref{fig_correlation}.
The measured 662 keV photopeak is found to be shifted by 15\% for GAGG(Ce) and by 22\% for LFS from a real photopeak.

\begin{equation}\label{eq4}
Q_{\rm real} = - Q_{\rm max}\ln{\left(\frac{Q_{\rm max}-Q_{\rm mes}}{Q_{\rm max}} \right)} 
\end{equation}

\begin{figure*}[htb]
	\begin{minipage}{0.5\hsize}
		\begin{center}
		\includegraphics[width=0.75\linewidth]{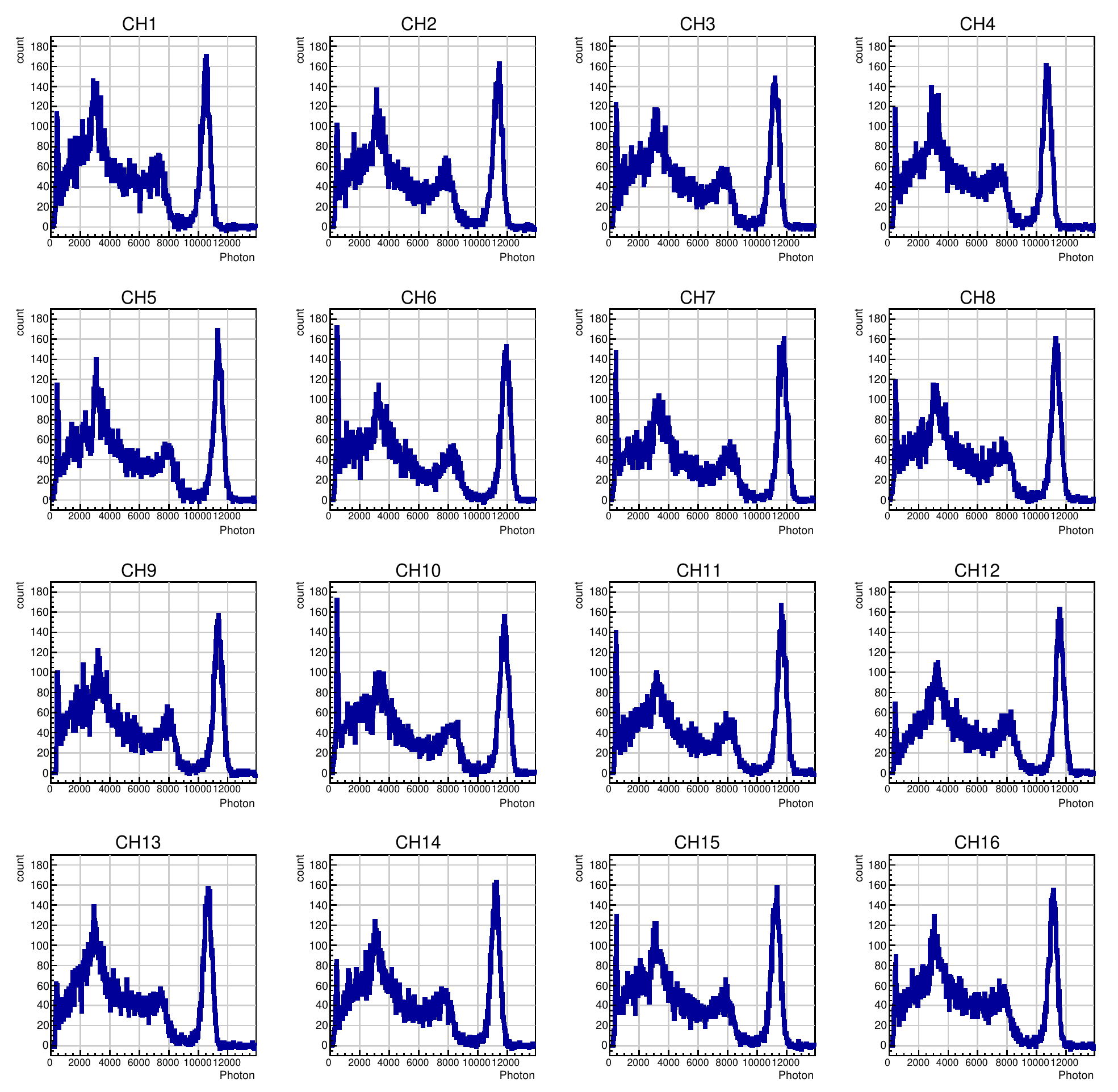}
		\end{center}
	\end{minipage}
	\begin{minipage}{0.5\hsize}
		\begin{center}
		\includegraphics[width=0.75\linewidth]{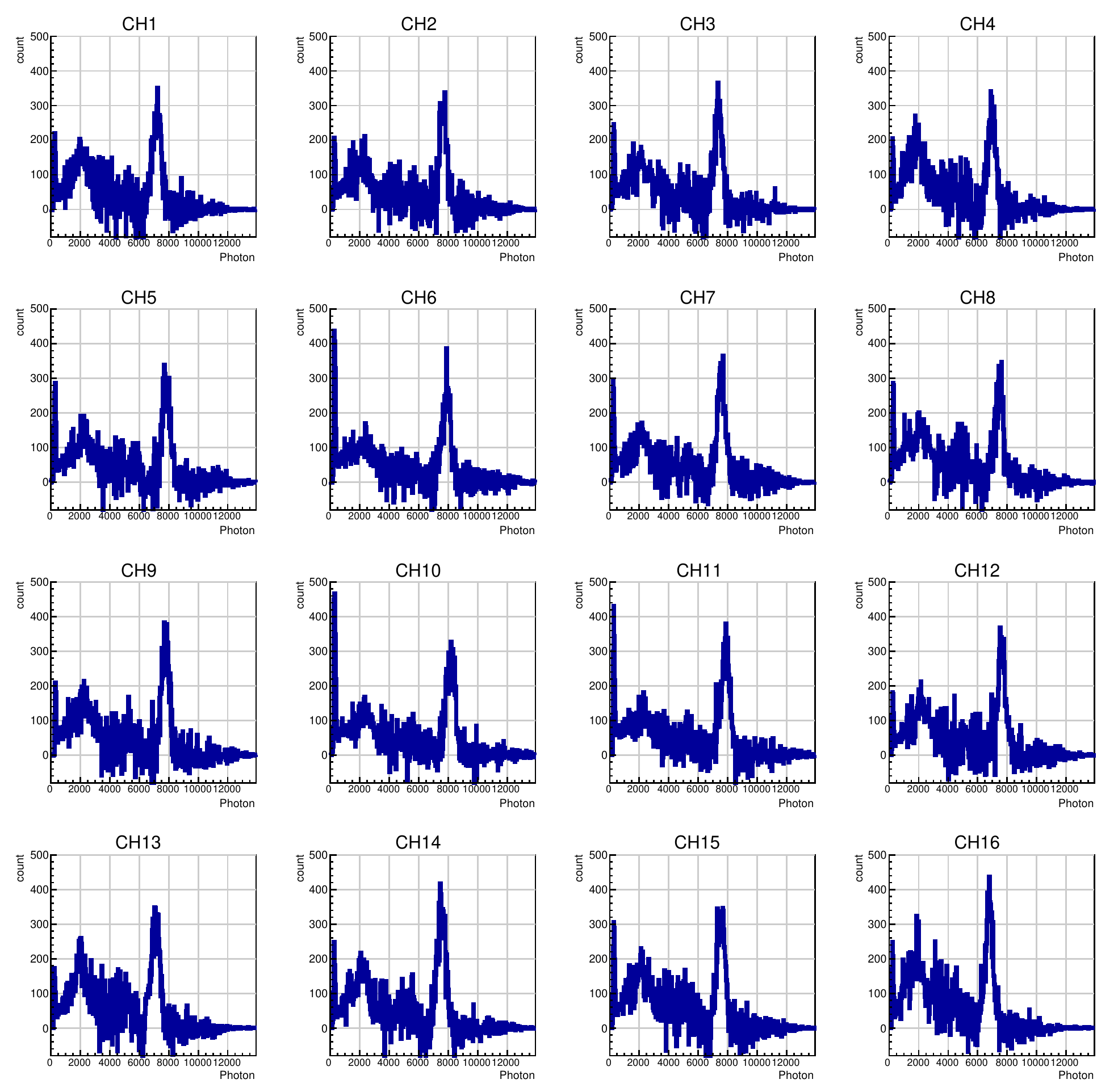}
		\end{center}
	\end{minipage}
	\caption{$^{137}\rm{Cs}$ spectra for each pixel of the GAGG(Ce) (left) and LFS (right) $4\times4$ array. The MPPC saturation and light leakage effects were corrected for these two spectra.}
	\label{array_sumspectra}
\end{figure*}

To collect all of the scintillation light scattered over the pixels,  
 we obtained charge $Q_{i,\rm sum}$ for a single-hit event in the pixel $i$
 by summing up the entire charge produced owing to light leakage in all pixels as follows.  
\begin{equation}\label{eq5}
Q_{i,\rm sum} = \sum_{j=1}^{16} Q_{j,\rm real} 
\end{equation}
Thus, the spectra derived for 16 individual pixels are shown in Figure \ref{array_sumspectra}. 
After light leakage corrections, the photopeak was increased by 52$\sim$94\% (GAGG(Ce)) and 
 21$\sim$32\% (LFS) compared with the original spectra at 20 $^\circ \rm{C}$. 

\begin{figure}[h]
\begin{center}
\includegraphics[width=0.85\linewidth,bb=0 0 518 408]{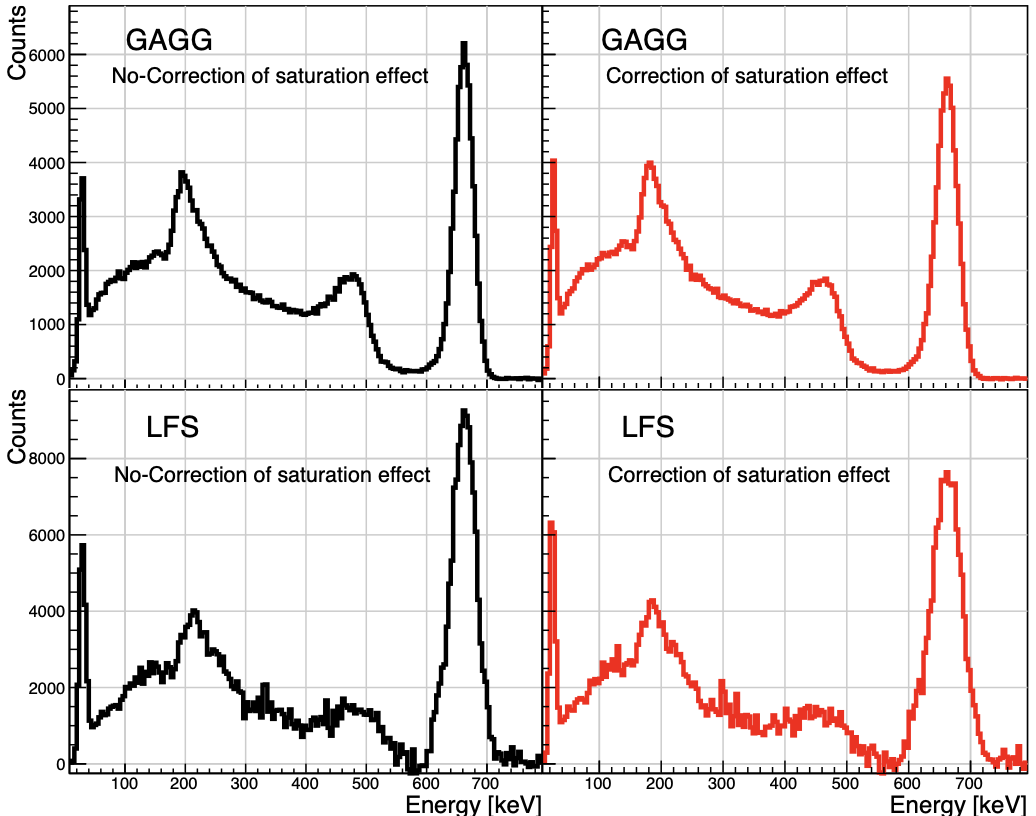}
\caption{16-channel summed $^{137}\rm{Cs}$ spectra for GAGG(Ce) (top) and LFS (bottom). The energy was calibrated using a 662 keV photopeak. The left (black) and right (red) panels show the spectra before and after the MPPC saturation correction, respectively.}
\label{sum_supe}
\end{center}
\end{figure}

Using 662 keV photopeak as an energy calibration, we further co-added 16-channel spectra after the gain correction. 
 Figure \ref{sum_supe} shows a comparison between without and with MPPC saturation correction. 
 Before the MPPC saturation effect correction, the full width half maximum (FWHM) energy resolution was 5.56$\pm$0.04\% for the GAGG(Ce) array and 7.11$\pm$0.13\% for the LFS array, whereas after the correction, both values significantly changed into 6.10$\pm$0.04\% for the GAGG(Ce) and 8.57$\pm$0.15\% for the LFS.
This means that the MPPC saturation effect causes a compression of the 662 keV photopeak, which looks apparently sharper than the real one.
As can be seen from Figure \ref{sum_supe}, we also found that the backscattered line at 184 keV was shifted to below 200 keV after the saturation correction.  
The energy resolution we obtained is a typical value (6\%-7\% for GAGG(Ce), and 8\%-9\% for LFS) in comparison with published references \cite{Yoneyama,Characterization_of_LFS3} even when we used an array and an ASIC readout system. 

The temperature dependence of the FWHM energy resolution is shown in Table \ref{resolution}. 
As the temperature increases, the energy resolution for both crystals worsens. 
The resolution is 5.75$\pm$0.04\% and 7.88$\pm$0.17\% at --20 $^{\circ}$C and 6.54$\pm$0.05\% and 8.38$\pm$0.15\% at 40 $^{\circ}$C for GAGG(Ce) and LFS, respectively. 
The thermal noise of the MPPC is also measured by applying the test pulse function to the ASIC, and its contribution to the energy resolution is negligible at only 0.02\% even at 40 $^{\circ}$C. 
The light yield at --20 $^{\circ}$C linearly increases by almost the same factor of approximately 1.20 for GAGG(Ce) and 1.23 for LFS from that measured at 40 $^{\circ}$C.
Assuming that the energy resolution is determined by the statistical fluctuation of the photo-electrons, the energy resolution should be proportional to the reciprocal of the square root of the light yield ($\propto 1/\sqrt{L.Y.}$). 
The expected value at --20 $^{\circ}$C is 5.98$\pm$0.04\% and 7.66$\pm$0.14\% using the 40 $^{\circ}$C values, which is consistent with the experimental results.   
Thus, we conclude that the energy resolution in the temperature range of --20 to 40 $^{\circ}$C is determined not by MPPC thermal noise but by the light yield of the scintillators.   

\setlength\intextsep{5pt}

% Conclusions
\section{Conclusions}
We constructed GAGG(Ce) and LFS 4$\times$4 scintillator arrays coupled to the Hamamatsu MPPC 4$\times$4 array, and evaluated the spectral performance and temperature dependence using an FWHM of a 662 keV peak by reading out the sensor signals using the ASIC IDE3380 provided by IDEAS.
By considering the MPPC saturation effect and applying the gain correction to each pixel based on the center of gravity information, we correctly obtained 16-channel summed spectra. The energy resolution obtained at 20 $^\circ$C is 6.10$\pm$0.04\% for GAGG(Ce) and 8.57$\pm$0.15\% for LFS.
This energy resolution is quite good even when we use an array of scintillators and an ASIC.
The energy resolution improves (e.g., 5.75$\pm$0.04\% for GAGG(Ce) and 7.88$\pm$0.17\% for LFS at --20 $^{\circ}$C) as the temperature decreases from 40 to --20 $^{\circ}$C. 
We found that the energy resolution of the MPPC+GAGG(Ce)/LFS array is determined only by the light yield of crystals with little effect from the thermal noise by the MPPCs.   
In terms of the internal background and spectral performance, we conclude that the GAGG(Ce) scintillator array is more suitable as an absorber of a Compton camera than the LFS array. 
This paper is the first evaluation test report of an array readout using an ASIC, and the sizes of the GAGG(Ce) crystals need to be optimized for future Compton camera application in space. 

\begin{table}[htb]
\begin{center}
\caption[]{Performance results for each scintillator array}\label{resolution}
\small
	\begin{tabular}{|c|c|c|c|c|}
		\hline
		Temperature & \multicolumn{2}{c|}{ FWHM energy resolution(\%) } \\
		\hline
		Degree ($^\circ$C ) & GAGG(Ce) & LFS \\
		\hline
		--20 & 5.75$\pm$0.04 & 7.88$\pm$0.17\\
		0 & 5.88$\pm$0.04 & 8.09$\pm$0.15 \\
		20 & 6.10$\pm$0.04 & 8.57$\pm$0.15 \\
		40 & 6.54$\pm$0.05 & 8.38$\pm$0.15 \\
		\hline
	\end{tabular}
\end{center}
\end{table}
 
\section{Acknowledgment}
We thank the anonymous referees for carefully reading our manuscript. 
This research was partially supported by the Toyoaki Scholarship Foundation and JSPS Grant-in-Aid for Scientific Research (KAKENHI) Grant No. JP18H03700 (KY).

\begin{comment}		
The author names and affiliations can be formatted in two ways:
\begin{enumerate}[(1)]
\item Group the authors per affiliation.
\item Use footnotes to indicate the affiliations.
\end{enumerate}
See the front matter of this document for examples. You are recommended to conform your choice to the journal you are submitting to.
\section{Bibliography styles}
There are various bibliography styles available. You can select that style of your choice in the preamble of this document. These styles are Elsevier styles based on standard styles like Harvard and Vancouver. Please use Bib\TeX\ to generate your bibliography and include DOIs whenever available.

Here are two sample references: \cite{Feynman1963118,Dirac1953888}.
\end{comment}

%\section*{References}
\bibliographystyle{elsarticle-num}

\end{document}